\documentclass[12pt]{article} 
\usepackage{amsmath}
\usepackage{amsfonts}
\usepackage{times}
\usepackage{epsf}
\def\({\left(}
\def\){\right)}
\def\[{\left[}
\def\]{\right]}
\def\non{ \nonumber }
\def\b{\beta}
\def\th{\theta}
\def\t{\text{t}}
\def\s{\text{s}}
\def\a{\alpha}
\def\si{\sigma}
\def\la{\lambda}
\def\e{\epsilon}

\begin{document} 
\begin{center}
{\bf Connecting lattice and relativistic models
 via conformal field theory.}
\end{center}
\phantom{a}
\vspace{1.5cm}

\centerline{H. E. Boos 
\footnote{on leave of absence from the Institute for High Energy Physics,
Protvino, 142284, Russia}}
\centerline{\it Institute for Solid State Physics}
\centerline{\it University of Tokyo, Kashiwa, Chiba 277-8581, Japan}

\phantom{a}

\vspace{0.5cm}

\phantom{a}

\centerline{V. E.  Korepin }
\centerline{\it C.N.~Yang Institute for Theoretical Physics}
\centerline{\it State University of New York at Stony Brook}
\centerline{\it Stony Brook, NY 11794--3840, USA}

\phantom{a}

\vspace{0.5cm}

\phantom{a}

\centerline{F.A. Smirnov
\footnote{Membre du CNRS}
}
\centerline{\it LPTHE, Tour 16, 1-er {\'e}tage, 4, pl. Jussieu}
\centerline{\it 75252, Paris Cedex 05, France}

\vspace{1cm}

\vskip2em
\begin{abstract}
\noindent
We consider the quantum group invariant XXZ-model.
In infrared limit it describes Conformal Field Theory (CFT)
with modified energy-momentum tensor. The correlation functions
are related to solutions of level -4 of  qKZ equations. We describe
these solutions relating them to level 0 solutions. We further consider
general matrix elements (form factors) containing local operators
and asymptotic states. We explain that the formulae for solutions
of qKZ equations suggest a decomposition of these matrix
elements with respect to states of corresponding CFT.
\end{abstract}

\newpage

\section
{Quantum group invariant XXZ-model.}

Let us recall some well known facts concerning XXZ-model and
its continuous limit.  Usually XXZ-model is considered as thermodynamic
limit of finite spin chain.
Consider the space $\(\mathbb{C}^2\)^{\otimes N}$. The finite spin chain in question
is described by the Hamiltonian:
\begin{align}
H_{XXZ}=\sum\limits _{k=1}^N(\si ^1_k\si _{k+1}^1+\si ^2_k\si _{k+1}^2+
\Delta\si ^3_k\si _{k+1}^3)\label{hxxz}
\end{align}
where the periodic boundary conditions are implied: $\si _{N+1}=\si _1$.
We consider the critical case $|\Delta |<1$ and parametrize it as follows:
$$\Delta =\cos \pi \nu$$
It is well-known that in the infrared limit the model
describes Conformal Field Theory (CFT) with $c=1$ and coupling constant
equal to $\nu$. The correlation functions in the thermodynamic limit
were found by Jimbo and Miwa \cite{jm}. 

It is equally matter of common knowledge that the model is closely
related to the R-matrix:
\begin{align}
R(\b, \nu)=
\left(
\begin{array}{cccc}
a(\b)&0&0&0\\
0&b(\b)&c(\b)&0\\
0&c(\b)&b(\b)&0\\
0&0&0&a(\b)
\end{array}
\right)
\label{R-m}
\end{align}
where
\begin{align}
&a(\b)= R_0(\b),\quad         b(\b)=  R_0(\b)\frac {\sinh \nu \b}
{\sinh \nu (\pi i-\b)}\non \\
&c(\b)=R_0(\b)\frac {\sinh \nu\pi i}{\sinh \nu(\pi i -\b)} \non
\\
& R_0(\b)=\exp
\left\{
i\int\limits _0^{\infty}
\frac {
\sin (\b k)\sinh\frac {\pi k(\nu -1)}{2\nu}
}
{
k\sinh\frac {\pi k}{2\nu}\cosh \frac{\pi k}{2  }
}
\right\}
\non
\end{align}
The coupling constant $\nu $ will be often omitted from $R(\b ,\nu)$.
The relation between R-matrix and XXZ-model is explained later.

>From the point of view of mathematics the R-matrix (\ref{R-m}) is
the R-matrix for two-dimensional evaluation representations of the
quantum affine algebra $U_q(\widehat{sl}_2)$. The latter algebra
contains two sub-algebras $U_q(sl_2)$. Let us perform a gauge
transformation with the R-matrix in order to make the invariance with
respect to one of them  transparent:
\begin{align} 
&\mathcal{R}(\b_1,\b_2,\nu)=
e^{\frac {\nu} 2\b _1\si ^3}\otimes e^{\frac {\nu} 2\b _2\si ^3}\ 
R(\b_1-\b _2,\nu)
\ e^{-\frac {\nu} 2\b _1\si ^3}\otimes e^{-\frac {\nu} 2\b _2\si ^3}=
\non\\&=
\frac   {R_0(\b_1-\b_2)}{2\sinh \nu (\pi i-\b_1+\b_2)}
\(e^{\nu (\b_1-\b _2)}
R_{21}^{-1}(q)-e^{\nu (\b_2-\b _1)}R_{12}(q)\)
\end{align}
where 
$$q=e^{2i\pi(\nu+1)}$$
Adding $1$ to $\nu$ is important since we will use fractional powers of $q$.
Here $R(q)$ is usual R-matrix for $U_q(sl_2)$:
$$
R_{12}(q)=
\left(
\begin{array}{cccc}
q^{\frac1 2}&0&0&0\\
0&1&q^{\frac1 2}-q^{-\frac1 2}&0\\
0&0&1&0\\
0&0&0&q^{\frac1 2}
\end{array}
\right)
$$
We want to use this quantum group symmetry. Unfortunately,
the Hamiltonian (\ref{hxxz}) is not invariant with respect to the action of
the quantum group which is represented  in the space $\(\mathbb{C}^2\)^{\otimes N}$
by 
\begin{align}
&S^3=\sum\limits _{k=1}^N \si ^3_k\non\\
&S^{\pm}=\sum\limits _{k=1}^N q^{-\frac{\si _1^{3}}4}\cdots q^{-\frac{\si_{k-1} ^{3}}4}
\si ^{\pm}_kq^{\frac{\si_{k+1} ^{3}}4}\cdots q^{\frac{\si_{N} ^{3}}4}\non
\end{align}
A solution of  this problem of quantum group invariance was
found by Pasquier and Saleur \cite{ps}. They proposed to consider another
integrable model on the finite lattice with Hamiltonian corresponding to
 open boundary conditions:
\begin{align}
H_{RXXZ}=\sum\limits _{k=1}^{N-1}(\si ^1_k\si _{k+1}^1+\si ^2_k\si _{k+1}^2+
\Delta\si ^3_k\si _{k+1}^3)
+i\sqrt{1-\Delta}\ (\si _1^3-\si _N^3)
\label{hrxxz}
\end{align}
This Hamiltonian is manifestly invariant under the action of quantum group
on the finite lattice. After the thermodynamic limit one obtains a model
with the same spectrum as original XXZ, but different scattering (this point will be
described later). The infrared limit corresponds to CFT with modified energy-momentum
tensor of central charge
$$c=1-\frac {6\nu ^2}{1-\nu}$$ 
especially interesting when $\nu$ is rational and additional restriction takes place.

In the present paper we shall consider RXXZ-model.
We shall propose formulae for correlators 
for this model showing their similarity with correlators for XXX-model.
The latter can be expressed in terms of values of Riemann 
zeta-function at odd natural arguments. We shall obtain an analogue
of this statement for RXXZ-model.

Let us say few words about hypothetic relation of XXZ and RXXZ models
in thermodynamic limit. The argument that this limit should not depend
on the boundary conditions must be dismissed in our situation since
we consider a critical model with long-range correlations. Still we would
expect that the following relation between two models in infinite volume exists.
The quantum group $U_q(sl_2)$ acts on infinite XXZ-model and commute 
with the Hamiltonian.
Consider a projector $\mathcal{P}$ on the invariant subspace.
We had XXZ-vacuum $|\text{vac}\rangle _{XXZ}$. 
We suppose that the 
RXXZ-model is obtained by
projection, in particular:
$$|\text{vac}\rangle _{RXXZ}=\mathcal{P}|\text{vac}\rangle _{XXZ}$$
The correlators in RXXZ-model are
$$
{\ }_{RXXZ}\langle \text{vac}|\mathcal{O}|\text{vac}\rangle _{RXXZ}=
{\ }_{XXZ}\langle \text{vac}|\mathcal{P}\mathcal{O}\mathcal{P}|\text{vac}\rangle _{XXZ}
$$
which can be interpreted in two ways: either as correlator in RXXZ-model
or as correlator of  $U_q(sl_2)$-invariant operator 
$\mathcal{P}\mathcal{O}\mathcal{P}$ in XXZ-model. 
This assumption explains the notation RXXZ  standing for Restricted XXZ-model.
So, we assume that in the lattice case a phenomenon close to the
one taking place in massive models occurs \cite{rs}. 

Let us explain in some more details the set of operators in XXZ model
for which we are able to calculate the correlators in simple form provided
the above reasoning holds. Under $\mathcal{O}$ we understand
some local operator of XXZ-chain, i.e. a product of several local
spins $\si ^a_k$, $a=1,2,3$. Under the above action of quantum group these
spins transform with respect to 3-dimensional adjoint representation. 
The projection $\mathcal{P}\mathcal{O}\mathcal{P}$ extracts
all the invariant operators, i.e. projects over the subspace of singlets
in the tensor product of 3-dimensional representations.

Let us explain more explicitly the relation between the the R-matrix and
XXZ, RXXZ Hamiltonians. Both of them can be constructed form the
transfer-matrix with different boundary conditions constructed via the
monodromy matrix:
$$ R_{01}(\la)R_{02}(\la )\cdots R_{0,N-1}(\la )R_{0,N}(\la )$$
In some cases it is very convenient to consider inhomogeneous model
for which the monodromy matrix contains a fragment:
$$R_{0k}(\la -\la _k)\cdots R_{0,k+n}(\la -\la _{k+n})$$
As we shall see many formulae become far more transparent for
inhomogeneous case. 

\section{QKZ on level -4 and correlators.}

The main result of Kyoto group \cite{jmmn,jm} is that the correlators in XXZ-model are
related to solutions of QKZ-equations \cite{SKyo,FR} on level -4. We formulate the
equations first and then explain the relation.
The equations for
the function $g(\b _1,\cdots,\b _{2n})\in \mathbb{C}^{\otimes 2n}$
are
\begin{align}
&R(\b_j -\b _{j+1})g(\b _1,\cdots ,\b _{j+1},\b _j,\cdots,\b _{2n})
=&\non\label{symm}\\
&=
\  \ g(\b _1,\cdots ,\b _{j},\b _{j+1},\cdots,\b _{2n})
\end{align}
\begin{align}
&g(\b _1,\cdots ,\b _{2n-1},\b _{2n}+2\pi i)
=
g(\b _{2n},\b _1,\cdots ,\b _{2n-1})
\label{Rie}
\end{align}
For application to correlators a particular solution is needed which satisfies
additional requirement:
\begin{align}
&g(\b _1,\cdots ,\b _{j},\b _{j+1},\cdots,\b _{2n})|_{\b _{j+1}=\b _j-\pi i}=
s_{j,j+1}\otimes g(\b _1,\cdots ,\b _{j-1},\b _{j+2},\cdots,\b _{2n})\label{norm}
\end{align}
where  $s_{j,j+1}$ is the vector $(\uparrow\downarrow)+(\downarrow\uparrow)$
in the tensor product of $j$-th and $(j+1)$-th spaces.

The relation of these equations to correlators is conjectured by
Jimbo and Miwa \cite{jm}. It cannot be proved for critical model
under consideration as it was done for the XXZ-model with $|q|<1$ in \cite{jmmn}.
However, later arguments based on Bethe Anzatz technique were 
proposed by Maillet and collaborators \cite{Maillet1,Maillet2} which
can be considered as a proof of Jimbo and Miwa conjecture.

Jimbo and Miwa find the solution needed \cite{jm} in the form:
\begin{align}
&g(\b _1,\cdots, \b _{2n})=\frac 1 {\sum e ^{\b_j}}
\prod\limits _{i<j}\zeta ^{-1}(\b _i-\b _j)
\int\limits _{-\infty}^{\infty} d\a _1\cdots \int\limits _{-\infty}^{\infty} 
d\a _{n-1}\prod\limits _{i,j}\varphi (\a _i ,\b _j,\nu)\non\\
&\times
\prod\limits _{i<j} \frac {A ^2_i-A^2_j}{a_i-a _jq}
\ D(a _1,\cdots a _{n-1}|b _1,\cdots ,b _{2n})\non 
\end{align}
where
$$\varphi (\a,\b,\nu)=\exp
\left\{-(1+\nu)\frac {\a+\b} 2
-2\int\limits _0^{\infty}
\frac {
\sin ^2(\frac {\a -\b} 2 k)\sinh\frac {\pi k(\nu +1)}{2\nu}
}
{
k\sinh\frac {\pi k}{2\nu}\sinh \pi  k
}
\right\}
$$
$\zeta(\b)$ is some complicated function, we shall not need it.
We use the notations:
$$ a_j=e^{2\nu\a _j},\quad b_j=e^{2\nu\b _j}, \quad A_j=e^{\a _j}\quad
B_j=e^{\b _j}$$
$D(a _1,\cdots a _{n-1}|b _1,\cdots ,b _{2n})$ is a Laurent polynomial of all its
variables taking values in $\mathbb{C}^{2n}$. We shall not use explicit
formula for this polynomial in the present paper.

For application to correlators in homogeneous XXZ-model
one has to specify: 
$$\b _1=\b _2 =\cdots =\b _n=-\frac {\pi i} 2$$
$$\b _{n+1}=\b _{n+2}=\cdots =\b _{2n}=\frac {\pi i}2$$
Then 
\begin{align}
&g\left(-\frac {\pi i} 2,\cdots, -\frac {\pi i} 2,\frac {\pi i} 2,\cdots ,\frac {\pi i} 2\right)
=\non\\&=
\int\limits _{-\infty}^{\infty} d\a _1\cdots \int\limits _{-\infty}^{\infty} 
d\a _{n-1}
\prod\limits _{i<j} \frac {A ^2_i-A^2_j}{a_i-a _jq}
\prod\limits _{i}\frac 1 {A _i +A_i^{-1}}
\ \ \widetilde{D}(a _1,\cdots ,a _{n-1})
\label{jm}
\end{align}
with some Laurent polynomial  $\widetilde{D}(a _1,\cdots ,a _{n-1})$.
The trouble with this integral is that it is essentially multi-fold one.
In our previous papers we have shown that the integrals can be simplified
and essentially reduced to products of one-fold ones in XXX case.
For the moment we cannot state the same for XXZ-model, but
we shall explain that the simplification can be done in RXXZ case.
Let us consider this in some more details.

According to the understanding of relation between XXZ and RXXZ models
explained in the Introduction we expect that the correlators for RXXZ
model are related to certain invariant under the quantum group 
solution of the same equations (\ref{Rie},\ref{norm}). In order to make the
quantum group symmetry transparent we make the transformation:
\begin{align}
\widehat{g}(\b _1,\cdots ,\b _{2n})=\exp \(\frac {\nu} 2\sum \b_j\si ^3_j\)
g(\b _1,\cdots ,\b _{2n})\non
\end{align}
With this notation the equations (\ref{Rie},\ref{norm}) take the form:
\begin{align}
&\mathcal{R}(\b_j,\b _{j+1})\widehat{g}(\b _1,\cdots ,\b _{j+1},\b _j,\cdots,\b _{2n})
=&\non\label{symm1}\\
&=
\  \ \widehat{g}(\b _1,\cdots ,\b _{j},\b _{j+1},\cdots,\b _{2n})
\end{align}
\begin{align}
&\widehat{g}(\b _1,\cdots ,\b _{2n-1},\b _{2n}+2\pi i)
=
-q^{-\frac 1 2 \si ^3_{2n}}\widehat{g}(\b _{2n},\b _1,\cdots ,\b _{2n-1})
\label{Rie1}
\end{align}
and
\begin{align}
&\widehat{g}(\b _1,\cdots ,\b _{j},\b _{j+1},\cdots,\b _{2n})
|_{\b _{j+1}=\b _j-\pi i}=
i\ \widehat{s}_{j,j+1}\otimes \widehat{g}(\b _1,\cdots ,\b _{j-1},\b _{j+2},\cdots,\b _{2n})\label{norm1}
\end{align}
where $\widehat{s}_{j,j+1}$ is the quantum group singlet in the tensor product 
of corresponding spaces:
$$q^{\frac 1 4}(\uparrow\downarrow)-q^{-\frac 1 4}(\downarrow\uparrow)$$

These equations respect the invariance under the quantum group. This
fact is obvious for the first and the third equations. To see this in the second
equation
one has to keep in mind that $q^{\frac 1 2 \si ^3}$ gives in two-dimensional
representation the element which realizes the square of antipode as inner
authomorphism. 

>From Jimbo-Miwa solution (\ref{jm}) one can obtain a solution to
(\ref{Rie1},\ref{norm1}) by projection on $U_q(sl_2)$- invariant subspace
which will suffer of the same problems related to denominators.
The main goal of this paper is to show that at least in this case corresponding
to RXXZ-model another form of solution is possible. 
\section
{QKZ on level 0.}

Consider the qKZ equations on level 0 which are the same as
two out of three basic equations (axioms) for
the form factors. We write these equations in $U_q(sl_2)$-invariant 
form which corresponds to form factors of RSG-model \cite{rs}.
Consider a co-vector
$\widehat{f}(\b _1,\cdots,\b _{2n}) \in \(\mathbb{C}^{\otimes 2n}\)^*$.
The equations are 
\begin{align}
&\widehat{f}(\b _1,\cdots ,\b _{j+1},\b _j,\cdots,\b _{2n})
=&\non\\
&=
\  \ \widehat{f}(\b _1,\cdots ,\b _{j},\b _{j+1},\cdots,\b _{2n})\mathcal{R}(\b_j-\b_{j+1})
\nonumber\\ &\ \non\\
&\widehat{f}(\b _1,\cdots ,\b _{2n-1},\b _{2n}+2\pi i)
=
-q^{-\frac 1 2 \si ^3_{2n}}\widehat{f}(\b _{2n},\b _1,\cdots ,\b _{2n-1})
\non
\end{align}
We need solution belonging to the singlet with respect to 
the action of $U_q(sl _2) $ subspace as has been explained in level -4
case. The application to form factors imposes additional requirement
which connects sectors with different number of particles:
\begin{align}
&2\pi i \text{res}_{\b _{2n}=\b _{2n-1}+\pi i}
\widehat{f}(\b _1\cdots,\b _{2n-2},\b _{2n-1},\b _{2n})=
\non\\ &=
\widehat{s}\ ^*_{2n-1,2n}\otimes \widehat{f}(\b _1\cdots,\b _{2n-2})
\(1-\mathcal{R}(\b_{2n-1}-\b_{1})\cdots \mathcal{R}(\b_{2n-1}-\b _{2n-2})\)
\label{a3}
\end{align}

The difference with level -4 case
 seems to be minor, but the formulae for
solutions are much nicer. Many solutions can be written 
which are counted sets of integers:
$\{k_1,\cdots ,k_{n-1}\}$
such that  $0\le k_1<,\cdots < k_{n-1}\le 2n-2$:
\begin{align}
&f^{\{k_1,\cdots ,k_{n-1}\}}(\b _1,\cdots, \b _{2n})=
\prod\limits _{i<j}\zeta (\b _i-\b _j)
\int\limits _{-\infty}^{\infty} d\a _1\cdots \int\limits _{-\infty}^{\infty} 
d\a _{n-1}\prod\limits _{i,j}\varphi (\a _i ,\b _j)\non\\
&\times
\text{det}\|A_i^{k_j}\|_{1\le i,j\le n-1}\
\ h(a _1,\cdots a _{n-1}|b _1,\cdots ,b _{2n})\prod _j a_jA_j\non 
\end{align}
where $h$ is skew-symmetric w.r. to $\a$'s polynomial. 
Notice that there are no
denominators mixing the integration variables in the integrant, so, effectively the integral is reduced to one-fold
integrals of the form:
\begin{align}
&\langle P\ |\ p\rangle=
\int\limits _{-\infty}^{\infty}\prod\limits _j\varphi (\a ,\b _j)\ P(A)\ p(a )
aAd\a\label{abint}
\end{align}
where $p(\a )$ and $P(A)$ are  polynomials. This is what we would like to have
for the correlators!

Again we do not describe explicitly the functions $h$ which take
values in $\(\mathbb{C}^{\otimes 2n}\)^*$. As has been said they
are skew-symmetric polynomials of $a's$ they are also rational
functions of $b's$ with simple poles at $b_i=qb_j$ only. But there is one
important property of $h$ which we need to mention.

First, the integral (\ref{abint}) is such that the degree of any
polynomial $s(a)$ can be reduced to $2n-2$ or less. 
For the polynomials of degree $\le 2n-2$ there is a basis (choice is not unique)
$$ s_j(\a ),\ j =-(n-1),\cdots , (n-1),\  \text{deg}(s_j)=j+(n-1)$$
with special properties described later. 
We shall not write down explicit formulae.
Then 
\begin{align}
&h(a _1,\cdots ,a _{n-1})=\sum\limits_{j_1\ne 0,\cdots j_{n-1}\ne 0}
\ h_{j_1,\cdots ,j_{n-1}}
\text{det}\|s_{j_p}(a _q)\|_{1\le p,q\le n-1}\non
\end{align}
and the skew-symmetric tensor
$h$ belongs to subspace of maximal irreducible representation of
symplectic group $Sp(2n-2)$ of dimension 
$$\text{dim}(\mathcal{H}_{\text{irreducible}})=\binom{2n-2}{n-1}-\binom{2n-2}{n-3}$$

Let 
$$J=1,\cdots ,\binom{2n-2}{n-1}-\binom{2n-2}{n-3}$$ 
Consider the basis $e^J$ in $\mathcal{H}_{\text{irreducible}}$ with components
$e^J_{j_1,\cdots j_{n-1}}$. Then we define
$h_J$ by
$$h_{j_1,\cdots ,j_{n-1}}=\sum\limits _{J}h_J\ e^J_{j_1,\cdots j_{n-1}}$$
$$\ $$
Recall that $h(a _1,\cdots ,a _{n-1})$ takes values in singlet subspace, so,
it has components
$$h^I(a _1,\cdots ,a _{n-1})$$
where $I$ counts basis of this subspace:
$$I=1,\cdots , \binom{2n}{n}-\binom{2n}{n-1}$$
Notice that
$$\binom{2n}{n}-\binom{2n}{n-1}=\binom{2n-2}{n-1}-\binom{2n-2}{n-3}$$
which means that there is a square matrix $h_J^I $ defined by
\begin{align}
&h^I(a _1,\cdots ,a _{n-1})=\sum\limits _{J}
h^I_J s^J(a _1,\cdots a _{n-1})\non
\end{align}
where 
$s_J(a _1,\cdots a _{n-1})$ are the following anti-symmetric polynomials:

\begin{align}
s^J(a _1,\cdots a _{n-1})=\sum\limits _{j_1,\cdots ,j_{n-1}}
e^J_{j_1,\cdots j_{n-1}}\text{det}\|s_{j_p}(a _q)\|_{1\le p,q\le n-1}
\non\end{align}
If we do not consider $s_0(a)$ the degrees of polynomials $P(A)$
can be reduced to $2n-3$ or less. We consider a special basis
$$ S_j(A),\quad |j|=1,\cdots ,(n-1), $$
$$\text{deg}(S_{-k})=2k-1, \ k=1,\cdots n-1,$$ 
$$\text{deg}(S_{k})=2k-2, \ k=1,\cdots n-1$$
which we do not describe explicitly, again.

The most important property of the integrals $\langle S_i\ |\ s_j\rangle$ is
deformed Riemann bilinear relation:
\begin{align}
&\sum\limits _{k=1}^{n-1}\left( \langle S_k\ |\ s_i\rangle \langle S_{-k}\ |\ s_j\rangle-
\langle S_k\ |\ s_j\rangle \langle S_{-k}\ |\ s_i\rangle\right)=\delta _{i, -j}\non\\
&\sum\limits _{k=1}^{n-1}\left( \langle S_i\ |\ s_k\rangle \langle S_j\ |\ s_{-k}\rangle-
\langle S_j\ |\ s_k\rangle \langle S_i\ |\ s_{-k}\rangle\right)=\delta _{i, -j}\non
\end{align}
These relations and properties of
$h(\a_1,\cdots ,\a _{n-1})$ imply that among $f^{\{k_1,\cdots ,k_{n-1}\}}$ only
$\text{dim} (\mathcal{H}_{\text{irreducible}})$ are linearly independent which
are span by action of $Sp(2n-2)$ on $\{1,3,\cdots, 2n-3\}$. 
The basis in this space is denoted by
\begin{align}
S_J(A _1,\cdots& , A _{n-1})=
\sum\limits _{j_1,\cdots ,j_{n-1}}
e_J^{j_1,\cdots ,j_{n-1}}\text{det}\|S_{j_p}(A_q)\|_{p,q=1,\cdots ,n-1}
\non\end{align}
The result is that the solutions are combined into square matrix 
(there is the same number of solutions as the dimension of space):
$$ F_I^J=P_I^{K}H_K^J$$
where $H_{KJ}$ is polynomial function of $\b _j$, the transcendental
dependence on $\b _j$ is hidden in the period matrix
$P_I^J$ which is defined as
$$P_I^J=\langle S^J\ |\ s_I\rangle$$
where the notations has obvious meaning:
$$ \langle P_1\wedge\cdots\wedge P_{n-1}\ |\ p_1\wedge\cdots\wedge p_{n-1}
\rangle=\text{det}\|\langle P_i \ |\ p_j\rangle \|_{1\le i,j \le n-1}$$

\section { New formula for level -4 from level 0.}

Recall that solutions to QKZ on level 0 are co-vectors while
solutions on level -4 are vectors. Consider the scalar product for two solutions:
$$f(\b _1,\cdots ,\b _{2n})g(\b _1,\cdots ,\b _{2n})$$
it is a quasi-constant (symmetric function of $e^{\b _j}$).
\newline
So, we can construct singlet solutions of QKZ on level -4 from those
on level 0.  Indeed we have square matrix $F$:
$$G=F^{-1}=H^{-1}P^{-1}$$
The matrix $H^{-1}$  is complicated but rational function of $\b _j$.
\newline
Due to deformed Riemann relation it is easy to invert $P$!
Indeed 
$$\(P^{-1}\)^I_J=\langle S ^I \ |\  s_J^{\dag}\rangle$$
where $s _J^{\dag}$ is obtained from $s _J$ replacing all 
$$s_j \longrightarrow \text{sgn} (j)s_{-j}$$ 
So, the transcendental part almost does not change, and we prove that the
new formula for solutions on level -4 is possible:
\begin{align}
&g^{\{k_1,\cdots ,k_{n-1}\}}(\b _1,\cdots, \b _{2n})=
\prod\limits _{i<j}\zeta (\b _i-\b _j)
\int\limits _{-\infty}^{\infty} d\a _1\cdots \int\limits _{-\infty}^{\infty} 
d\a _{n-1}\prod\limits _{i,j}\varphi (\a _i ,\b _j)\non\\
&\times
\text{det}\|A_i^{k_j}\|_{1\le i,j\le n-1}\
\ \tilde{h}(a _1,\cdots a _{n-1}|b _1,\cdots ,b _{2n})\prod _j a_jA_j\non 
\end{align}
where $\tilde{h}$ are skew-symmetric w.r. to $a _i$ polynomials.
Actually, they are polynomials in $b _j$ as well. The proof is based on the
following calculation:
$$ H^{-1}=H^*\( H H^* \)^{-1}$$
The operator $H H^*$ is nicer than $H$ itself because it acts from 
$\mathcal{H}_{\text{irreducible}}$ to itself. We were able to calculate its determinant:
\begin{align}
&\text{det}(H
H^{*})
=Const
\(\prod\limits _{i,j}(b _i-qb _j)\)^{-\(\binom{2n-4}{n-2}-
\binom{2n-4}{n-4}\)}\non
\end{align}
Also the rank of the residue of $H$ at $b _j=q\b _i$ equals
the dimension of singlet subspace in
$\mathbb{C}^{\otimes (2n-2)}$.
\section
{Cohomological meaning of new formula.}

"Classical" limit:
$ \nu\to 0$ and $\b _j$ are rescaled in such a way that 
$b _j $ are finite.
In this limit
\begin{align}
&\langle P\ |\ p\rangle=
\int\limits _{-\infty}^{\infty}\prod\limits _j\varphi (\a ,\b _j)\ P(A)\ p(a )
d\a\non\to\non\\&
\to \int\limits _{\gamma}\frac {p(a)} cda\non
\end{align}
where the hyper-elliptic surface $X$ is defined by
$$c^2=\prod (a-b_j),$$
The genus equals $n-1$.
The contour $\gamma$ is defined by $P$. In particular,
$$S_{-k}\leftrightarrow\text{b}_k, \quad S_{k}\leftrightarrow \text{a}_k$$
Consider
$$\text{Symm}(X^{n-1})$$
the points on this variety are  divisors:
$$\{P_1,\cdots , P_{n-1}\}\qquad P_j=\{a_j,c_j\}\in X$$
Consider the non-compact variety
$$\text{Symm}(X^{n-1})-D$$
where 
$$D=\{\{P_1,\cdots , P_{n-1}\}|P_j=\infty ^{\pm}, P_i=\sigma(P_j)\}$$
This is an affine variety isomorphic to affine Jacobian.

The integrant of the classical limit of invariant part of Jimbo-Miwa
solution (\ref{jm}) gives a $(n-1)$-differential form (maximal dimension) on
$$\text{Symm}(X^{n-1})-D$$
of the kind:
\begin{align}
&\Omega =\frac {F(a_1,c_1,\cdots , a_{n-1},c_{n-1})}{\prod _{i<j}(a_i-a_j)}
\ \frac{da_1}{c_1}\wedge\cdots\wedge\frac{da_{n-1}}{c_{n-1}}\non
\end{align}
where the polynomial $F(a_1,c_1,\cdots , a_{n-1},c_{n-1})$ vanishes
when $a_i=a_j$ and $c_i=c_j$.
The question arises concerning cohomologies.
\newline
{\bf Theorem} (A. Nakayashiki) The elements of $H^{(n-1)} $ can be
realized as 
\begin{align}
&\Omega _{k_1,\cdots ,k_{n-1}}=
\text{det}\| a_p^{k_q}\|_{p,q=1,\cdots ,n-1}
\quad\frac{da_1}{c_1}\wedge\cdots\wedge\frac{da_{n-1}}{c_{n-1}}\non
\end{align}
where $k_q=0,\cdots ,2n-2$.\newline
{\it Remark.} Actually  some of these forms are linearly dependent 
(mod exact forms), we do not describe all details.

\section
{ Back to correlattors.}

Comparing with  Jimbo-Miwa solution 
one makes sure that the solution needed for correlators is 
$$ g^{\{0,2,4,\cdots ,2n-4\}}$$
so, it corresponds to "a-cycles".
We need to put
\begin{align}
\b_k=\la _k-\frac{\pi i}{2}+i\delta _k,\quad
\b_{2n-k+1}=\la _k+\frac{\pi i}{2}-i\delta _k 
\non
\end{align}
and to take he limit $\delta _k\to 0$. The calculation of integrals
is similar to XXX case, 
the result can be expressed in terms of the function:
\begin{align}
&\chi (\a)=\frac d {d\a}\(\log\frac{\varphi (\a -\frac{\pi i} 2)}
{\varphi (\a +\frac{\pi i} 2)}\)=i\int\limits _0^{\infty}
\frac {
\cos (\a k)\sinh\frac {\pi k(\nu -1)}{2\nu}
}
{
\sinh\frac {\pi k}{2\nu}\cosh \frac{\pi k}{2  }
}dk=\non\\
&=i\sum\limits _{m=0}^{\infty}\a ^{2m}\ \frac {(-1)^m} {(2m)!}\int\limits _0^{\infty}
\frac {
k^{2m}\sinh\frac {\pi k(\nu -1)}{2\nu}
}
{
\sinh\frac {\pi k}{2\nu}\cosh \frac{\pi k}{2  }}dk
\non
\end{align}
Finally for the correlator in inhomogeneous case:
\begin{align}
&g(\la_1 -\frac {\pi i} 2\cdots\la_n -\frac {\pi i} 2,\la _{n}+\frac {\pi i} 2
\cdots \la _{1}+\frac {\pi i} 2)
^{\e _1\cdots\e _{n} \e _{n+1}\cdots\e _{2n}}=
\non\\&=
\sum\limits _{m=0}^{\[\frac n 2\]}
\sum\limits _{k_1,\cdots ,k _{2m}}
Q_{k_1k_2\cdots  k_{2m-1}k _{2m}}
^{\e _1\cdots\e _{n} \e _{n+1}\cdots\e _{2n}}
(\la _1,\cdots, \la _n)
\non\\&\times
\chi(\la _{k_1}-\la _{k_2})\cdots \chi(\la _{k_{2m-1}}-\la _{k_{2m}})
\non
\end{align}

\section
{\bf  General matrix elements.}

When we pass to description of XXZ-model in terms of particles
a common phenomenon known nowadays as "modular double" \cite{fad}
occurs. The essence of this phenomenon is that another quantum group
with dual $q$ enters the game. In a sense RXXZ model is invariant
with respect to "modular double" which is quite non-trivial, and 
not completely understood,  combination of two quantum groups. 
The particle description of the model is as follows.

For coupling constants
not very far from $0$ the spectrum
of the model contains one particle (magnon).  
This particle is parametrized by rapidity $\th$ carrying momentum and energy:
$$ p(\th)=\log \tanh \frac 1 2 \left(\th -\frac {\pi i} 2 \right), \quad
e(\th)=\frac {dp(\th)}{d\th}$$
The particle has internal degrees living in isotopic space $\mathbb{C}^2$.
The S-matrix is given by
$$S(\th _1	-\th _2)=R(\th _1	-\th _2, \ \textstyle{\frac {\nu}{1-\nu}})$$
This is where the second quantum group appears. The RXXZ model is
invariant under the action of of two quantum groups:
$$U_q(sl _2)\ , \ U_{\widetilde{q}}(sl _2), \quad \text{with} \quad q=e^{2\pi i (\nu+1)}
\ ,
\ \widetilde{q}=e^{\frac { 2\pi i }{1-\nu}}$$
For the asymptotic states it means that they must be taken as
invariant under the action of the second quantum group. All that is
familiar from consideration of massive models and its restrictions \cite{rs}.

Consider the matrix elements
$${\ }_{RXXZ}\langle \text{vac}\ |\ \mathcal{O}\ |\ \th _1,\cdots ,\th _n\ \rangle_{RXXZ}$$
where $\mathcal{O}$ is some operator of the type
$$E_{\e _1}^{\e_1'}\cdots E_{\e _n}^{\e_n'}$$
It can be obtained from "Kyoto generalization"  which is the function
$$\widehat{ f}(\b _1,\cdots ,\b_{2n}, \th _1,\cdots ,\th _{2m})\in \mathbb{C}^{\otimes 2n}
\otimes \left(\mathbb{C}^*\right)^{\otimes 2m}$$
which satisfies level -4 qKZ with R-matrix $\mathcal{R}(\cdot , \nu)$ (denoted by $\mathcal{R}(\cdot)$) with respect to
$\b $'s and level 0 qKZ with 
gauge transformed S-matrix 
$\mathcal{R}(\cdot, \ \textstyle{\frac {\nu}{1-\nu}})$
(denoted by $\mathcal{S}(\cdot )$)
with respect to $\th$'s. Actually, both equations are slightly modified. In addition it must satisfy the following
normalization conditions. All together we have:
\begin{align}
&\mathcal{R}(\b _{j+1} -\b_j)
\widehat{f}(\b _1,\cdots ,\b _{j+1},\b _j,\cdots,\b _{2n},\th _1,\cdots ,\th _{2m})
=\label{s1}\\
&=
\  \ \widehat{f}(\b _1,\cdots ,\b _{j},\b _{j+1},\cdots,\b _{2n},\th _1,\cdots ,\th _{2m})\non
\end{align}
\begin{align}
&\widehat{f}(\b _1,\cdots ,\b _{2n-1},\b _{2n}+2\pi i,\th _1,\cdots ,\th _{2m})
=\label{r1}\\&=-\prod\limits _{j=1}^{2m}\tanh \frac 1 2 \(\b_{2n}-\th _j+\frac {\pi i} 2\)
q^{\frac 1 2 \si ^3_{2n}}\widehat{f}(\b _{2n},\b _1,\cdots ,\b _{2n-1},\th _1,\cdots ,\th _{2m})
\non
\end{align}
\begin{align}
&\widehat{f}(\b _1,\cdots ,\b _{2n-2},\b _{2n-1},\b _{2n},\th _1,\cdots ,\th _{2m})|
_{\b _{2n}=\b _{2n-1}+\pi i}=\label{n1}\\ &=
\widehat{s}_{2n-1,2n}\otimes 
\widehat{f}(\b _1,\cdots ,\b _{2n-2},\th _1,\cdots ,\th _{2m})\non
\end{align}
\begin{align}
&\widehat{f}(\b _1,\cdots,\b _{2n},\th _1, \cdots ,\th _{j+1},\th _j, \cdots, \th _{2m})
=\label{s2}\\
&=
\  \ \widehat{f}(\b _1,\cdots,\b _{2n},
\th _1, \cdots,\th _j ,\th _{j+1},\cdots, \th _{2m})\mathcal{S}(\th _j-\th _{j+1})
\nonumber\\ &\ \non
\end{align}
\begin{align}
&\widehat{f}(\b _1,\cdots,\b _{2n},\th _1,\cdots ,\th _{2m-1},\th _{2m}+2\pi i)
=\label{r2}\\&=
-\prod\limits _{j=1}^{2n}\tanh \frac 1 2 \(\th_{2m}-\b _j+\frac {\pi i} 2\)
\widehat{f}(\b _1,\cdots,\b _{2n},\th _{2m},\th _1,\cdots ,\th _{2m-1})
q^{-\frac 1 2 \si ^3_{2m}}
\non
\end{align}
\begin{align}
&2\pi i \text{res}_{\th _{2m}=\th _{2m-1}+\pi i}
\widehat{f}(\b _1,\cdots,\b _{2n},\th _1\cdots,\th _{2m-2},\th _{2m-1},\th _{2m})=
\non\\ &=
\widehat{s}{\ }^*_{2m-1,2m}
\otimes \widehat{f}(\b _1,\cdots,\b _{2n},\th _1\cdots,\th _{2m-2})
\label{n2}\\
&\times \(1-
\prod\limits _{j=1}^{2n}\tanh \frac 1 2 \(\th_{2m-1}-\b _j+\frac {\pi i} 2\)
\mathcal{S}(\th_{2m-1}-
\th_{1})\cdots \mathcal{S}(\th_{2m-1}-\th _{2m-2})\)
\non
\end{align}
The equations (\ref{s1},\ref{r1},\ref{s2},\ref{r2}) are slightly different
from respectively level -4 and level 0 qKZ equations because of
multipliers containing $\tanh $'s. This difference, however, is easily
taken care of by multiplier
$$\prod\limits _{i=1}^{2n}\prod\limits _{j=1}^{2m}\psi (\b _i,\th _j)$$
where the  function 
$$\psi (\b,\th )= 2^{-\frac 3 4}\exp \(-\frac {\b +\th } 4 -\int\limits _{0}^{\infty}
\frac {\sin ^2\frac 1 2 (\b -\th +\pi i)k+\sinh ^2 \frac {\pi k} 2}
{k\sinh \pi k \cosh \frac {\pi k} 2}dk\)$$
satisfies the equations:
\begin{align}
&\psi (\b, \th +2\pi i)=\tanh \frac 1 2 (\th-\b +\frac {\pi i}2)\psi (\b, \th)\non\\
&\psi (\b,\th)\psi (\b,\th +\pi i)=\frac 1 {e^{\b}-i\e ^{\th}}\non
\end{align}

For  the function 
$ f(\b _1,\cdots ,\b_{2n}, \th _1,\cdots ,\th _{2m})$ in XXZ-model Jimbo-Miwa give
a formula of the following kind:
\begin{align}
& f(\b _1,\cdots ,\b_{2n}, \th _1,\cdots ,\th _{2m})=
\prod\limits _{i<j}\zeta (\th _i-\th _j,\textstyle{\frac {\nu}{1-\nu}})
\prod\limits _{i<j}\zeta ^{-1}(\b _i-\b _j,\nu)
\prod\limits _{i,j}\psi (\b _i, \tau _j)
\non\\&\ \non\\&\times
\int\limits _{-\infty}^{\infty}d\a_1\cdots
\int\limits _{-\infty}^{\infty}d\a _{n}
\int\limits _{-\infty}^{\infty}d\si _1\cdots
\int\limits _{-\infty}^{\infty}d \si _m
\prod \varphi (\a _i-\b_j,\nu)
\prod \varphi (\si _i-\th_j, \textstyle{\frac {\nu}{1-\nu}}      )\non\\&\ \non\\ &\times
\prod\limits _{i<j}\frac {A^2_i-A^2_j}{a_i-qa_j}\ \prod\limits _{i<j}
\frac {S^2_i-S^2_j}{\s _i-\widetilde{q}\s_j}
\ \prod \frac 1 {A^2_i-S^2_j}\non\\&\ \non\\ &\times
D(a_1,\cdots ,a_{n}|b_1,\cdots ,b_{2n})F(\s_1,\cdots ,\s _{m}|\t_1, \cdots \t_{2m})
\non
\end{align}
where we use the notations:
\begin{align}
&a _j=e^{2\nu \a _j},\quad b_j=e^{2\nu \b _j}, 
\quad A_j=e^{\a _j}, \quad B_j=e^{\b _j}\non\\
&\s _j=e ^{\frac {2\nu}{1-\nu} \si _j},\quad  \t _j =e ^{\frac {2\nu}{1-\nu} \th _j},
\quad S_j =e^{\si _j},\quad , T_j=e^{\th _j}\non
\end{align}
The functions $D$, $F$ are polynomials of their variables.
For us the main problem with this formula is in denominators.
Here we are concerned not only about the denominators 
$a_i-qa_j$ and $\s _i-\widetilde{q}\s_j$ which are unpleasant for technical reasons as
explained above. Our main trouble is in the denominators $A^2_i-S^2_j$
because due to certain physical intuition we would expect another
kind of formula. Let us explain the point. 

At this point it would be more clear to talk about lattice SOS-model
instead of RXXZ-model. These two models are equivalent due to
usual Onzager relation between 2D classical statistical physics
and 1D quantum mechanics. The advantage of the lattice
model is due to the fact that it allows  intuitively clear relation to
Euclidian Quantum Field Theory. Our physical intuition about
general matrix element is based on the following picture:
\vskip 1cm
\hskip 2.6cm  
\epsffile{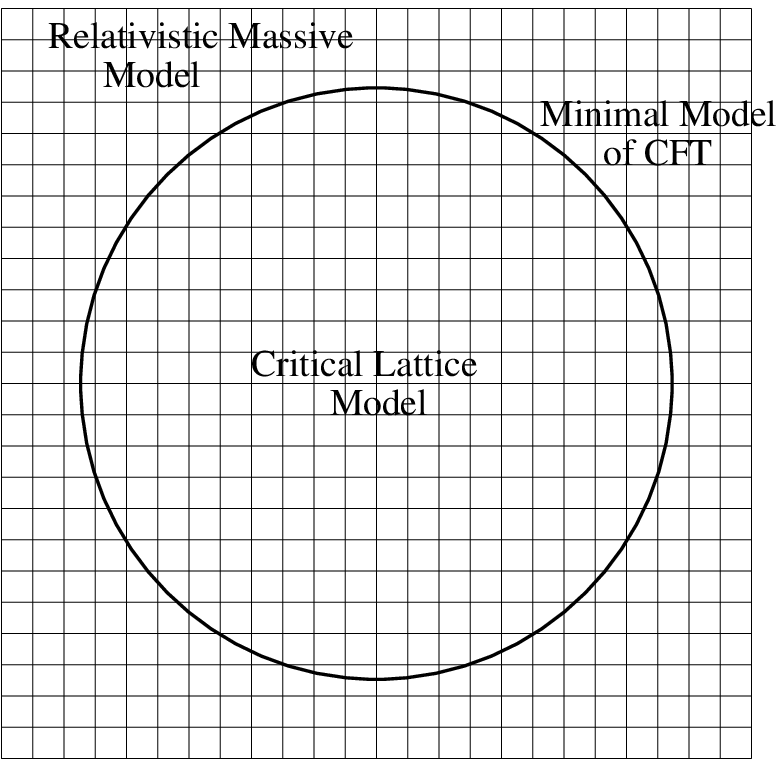}
\vskip 0.2cm
\noindent
Let us give some explanations. Suppose we consider instead of 
critical model  SOS-model out of criticality corresponding to
elliptic R-matrix. Suppose further that we are very
close to the critical temperature. Then microscopically we have
already critical lattice SOS-model. On the scales much bigger than
the lattice size but much less than the correlation length we
have massless relativistic field theory which is nothing
but CFT with the central charge
$$c=1-\frac {6\nu ^2}{\nu -1}$$
Finally on the scales of the order of correlation length
we have massive relativistic field theory which is RSG-model
with the coupling constant $\frac {1-\nu }{\nu}$. The role of CFT is clear:
it describes infrared limit of the lattice model and ultraviolet limit
of massive model. The local operators of the massive model are
counted by the states of CFT. These local operators are
described by form factors in asymptotic states description.
On the other hand one should be able to consider the lattice
critical model with boundary conditions corresponding to different
states of CFT.  That is why we expect the following kind
of formula for general matrix element:
\begin{align}
{\ }_{RXXZ}\langle \text{vac}|\mathcal{O}|\th _1,\cdots ,\th _{2m}\rangle=
\sum\limits _{\Psi} {\ }_{RXXZ}\langle \text{vac}|\mathcal{O} |\Psi\rangle
\langle \Psi {\ }|\th _1,\cdots ,\th _{2m}\rangle\label{dec}
\end{align}
where $\Psi$ are states of CFT, $\langle \Psi {\ }|\th _1,\cdots ,\th _{2m}\rangle$
are form factors of local operator corresponding to $\Psi$ in RSG-model,
${\ }_{RXXZ}\langle \text{vac}|\mathcal{O} |\Psi\rangle $ are correlators
of local operator $\mathcal{O}$ in the lattice model (of usual kind 
$E_{\epsilon _1}^{\epsilon '_1}\cdots E_{\epsilon _n}^{\epsilon '_n}$).
The latter object requires more careful definition, we hope to
return to it in feature. 

Notice that the formula (\ref{dec}) is in nice correspondence with the system
of equations (\ref{s1}, \ref{r1}, \ref{n1}, \ref{s2}, \ref{r2}, \ref{n2}) because
passing from the Kyoto generalize correlator to the usual one
we put $\b _{j}=\b _{2n-j+1}+\pi i$, so, the $th $ in equations with respect
to $th _j$ cancel, and we get usual Form Factor Axioms.  In our case
of RSG-model a complete set of solutions to these axioms is known 
\cite{count,bbs}, so, a formula of the kind (\ref{dec}) must hold.

So, there must be a formula of the type:
\begin{align}
& f(\b _1,\cdots ,\b_{2n}, \th _1,\cdots ,\th _{2m})=
\prod\limits _{i<j}\zeta (\th _i-\th _j,\textstyle{\frac {\nu}{1-\nu}})
\prod\limits _{i<j}\zeta ^{-1}(\b _i-\b _j,\nu)
\prod\limits _{i,j}\psi (\b _i, \tau _j)\non\\&\ \non\\&\times
\int\limits _{-\infty}^{\infty}d\a_1\cdots
\int\limits _{-\infty}^{\infty}d\a _{n}
\int\limits _{-\infty}^{\infty}d\si _1\cdots
\int\limits _{-\infty}^{\infty}d \si _m
\prod \varphi (\a _i-\b_j,\nu)
\prod \varphi (\si _i-\th_j, \textstyle{\frac {\nu}{1-\nu}}      )\non\\&\ \non\\ &\times
M(A_1,\cdots ,A_{n-1}|T_1,\cdots ,T_{m-1})
\non\\ &\times
\widetilde{h}(a_1,\cdots ,a_{n}|b_1,\cdots ,b_{2n})h(\s_1,\cdots ,\s _{m}|\t_1, \cdots \t_{2m})
\label{pol}
\end{align}
where $M(A_1,\cdots ,A_{n-1}|S_1,\cdots ,S_{m-1})$ is skew-symmetric with respect 
to 
\newline$A_1,\cdots ,A_{n-1}$ and $T_1,\cdots ,T_{m-1}$ polynomial which
depends on $B_j$, $S_j$ as on parameters. This polynomial must
satisfy certain equations in order that the relations (\ref{n1}, \ref{n2}) hold.
We do not write down explicitly these bulky equations, but fortunately
they coincide with equations for similar polynomials for quite
different problem which is the calculation of form factors for massless
flows \cite{ms}. The solution to these equations is not unique, but
there is a "minimal" one which has minimal possible degree with 
respect to variables $A_j$ and $S_j$. Our conjecture is that this is
the solution we need. It satisfies all simple checks that we were able
to carry on. Denote the sets $S=\{1,\cdots ,2n\}$, $S'=\{1,\cdots ,2m\}$.
The polynomial is:
\begin{align}
&M(A_1,\cdots ,A_{n-1}|S_1,\cdots ,S_{m-1})=\non\\&=
\prod\limits _{i<j}(A_i-A_j)\prod\limits _{i<j}(S_i-S_j)
\prod\limits _{j=1}^{2n}B_j\prod\limits _{j=1}^{n-1}A_j\non\\&\times
\sum\limits   _{T\subset S\atop\#T=n-1}\sum\limits _{T'\subset S'\atop\#T'=m-1}
\prod\limits _{j\in T}B_j\prod\limits _{i=1}^{n-1}\prod\limits _{j\in T}(A_i+iB_j)
\prod\limits _{i=1}^{m-1}\prod\limits _{j\in T'}(S_i+iT_j)\non\\
&\times \prod\limits _{i,j\in S\backslash T\atop i<j}(B_i+B_j)
\prod\limits _{i,j\in S'\backslash T'\atop i<j}(T_i+T_j)
\prod\limits _{i\in T\atop j\in S\backslash T}\frac 1 {B_i-B_j}
\prod\limits _{i\in T'\atop j\in S'\backslash T'}\frac 1 {T_i-T_j}\non\\ &\times
\prod\limits _{i\in T\atop j\in S'\backslash T'}(B_i+iT_j)
\prod\limits _{i\in T'\atop j\in S\backslash T}(T_i+iB_j)
X_{T,T'}(B_1,\cdots ,B_{2n}|T_1,\cdots ,B_{2m})\non
\end{align}
where 
\begin{align}
&X_{T,T'}(B_1,\cdots ,B_{2n}|T_1,\cdots ,B_{2m})=\non\\
&=\sum\limits _{i_1,i_2\in S\backslash T}\prod\limits _{p=1}^2
\(\frac {\prod _{j\in T}(B_{i_p}+B_j) \prod _{j\in T'}(B_{i_p}+iT_j)}    
{\prod _{j\in S'\backslash T'\backslash \{i_1,i_2\}}(B_{i_p}-B_j)
\prod _{j\in S'\backslash T'}(B_{i_p}-iT_j)}\)\non
\end{align}
Obviously, the formula (\ref{pol}) is in agreement with the intuitive
formula (\ref{dec}). After specialization $\b _k =\la _k +\frac {\pi i}2$,
$\b _{2n-k+1} =\la _k -\frac {\pi i}2$ (\ref{pol}) will turn into a sum
of form factors of RSG-model with coefficients constructed via
the functions $\chi (\la _i-\la _j)$ which correspond to correlators
of RXXZ-model with boundary conditions. The identification of
RSG-form factors with operators counted by CFT is known at least
to some extent \cite{bbs,nak,jm1}. So, it should be possible to 
make the correspondence between (\ref{pol}) and (\ref{dec}) more
explicit, but this problem goes beyond the scope of the present paper.

\noindent{\bf Acknowledgments.} HEB would like to thank 
Masahiro Shiroishi, Pavel Pyatov and Minoru Takahashi 
for useful discussions.
This research  has been supported by 
the following grants: the Russian Foundation of Basic Research
under grant \# 01--01--00201, by INTAS under grants \#00-00055 and 
\# 00-00561
and by EC network "EUCLID",
contract number HPRN-CT-2002-00325. HEB would also 
like to thank the administration
of the ISSP of Tokyo University for hospitality and  
perfect work conditions. The research of VEK was supported by NSF Grant
PHY- 0354683. This paper is based on the talk given by FAS at 
"Infinite Dimensional Algebras  and Quantum Integrable Systems"
(Faro, Portugal, July 21-25, 2003), FAS is grateful to organisers for their
kind hospitality.

\end{document}